# A New Property of Hamilton Graphs

Heping JIANG

Rm. 702, Building 12, Luomashi Avenue, Xi Cheng district,
Beijing, China, zip code: 100052

E-mail address: jjhpjhp@gmail.com

**Abstract.** A Hamilton cycle is a cycle containing every vertex of a graph. A graph is called Hamiltonian if it contains a Hamilton cycle. The Hamilton cycle problem is to find the sufficient and necessary condition that a graph is Hamiltonian. In this paper, we give out some new kind of definitions of the subgraphs and determine the Hamiltoncity of edges according to the existence of the subgraphs in a graph, and then obtain a new property of Hamilton graphs as being a necessary and sufficient condition characterized in the connectivity of the subgraph that induced from the cycle structure of a given graph.

## 1. Introduction

Graphs considered throughout this paper are finite, undirected and simple connected graphs. A Hamilton cycle is a cycle containing every vertex of a graph. A graph is called Hamiltonian if it contains a Hamilton cycle. The Hamilton cycle problem is to find the necessary and sufficient condition that a graph is Hamiltonian. Since Eulerian graphs have a simple characterization, the study of the relationship between Hamilton graphs and degree of vertices became a classical method and received numerous results [1]. In the meantime, different from the classical method, a kind of quantitative analysis, S. Goodman and S. Hedetniemi investigated the relationship between subgraphs and the Hamilton graphs and developed a qualitative analysis [2]. From then on the subgraphs have been wildly studied [1] [3]. However, the Hamilton cycle problem is still an open problem. In this paper, using the qualitative analysis, we give out some new kind of definitions of the subgraphs and determine the Hamiltoncity of edges according to the existence of the subgraphs in a graph, and then obtain a new property of Hamilton graphs as being a necessary and sufficient condition characterized in the connectivity of the subgraph that induced from the cycle structure of a given graph. Based on this property, we will show a polynomial algorithm for finding a Hamilton cycle in a graph in the other paper.

Let $G$ be a graph with vertex set $V$ and edge set $E$. We use combination to represent the overlapping of two graphs (or cycles) based on common vertices and edges. A cycle-set is the combination of $V-E+1$ cycles of G such that it covers all vertices and edges of G. The union of all elements in a cycle-set is the graph itself. $F_s$ is a spanning cycle-set, denotes a set of minimum number of cycles covering all vertices in G. In a cycle-set of G, every node corresponding to the vertex in G implies the overlap of vertices of the cycles passing through. So does an edge in G. For brevity, we reserve the word "vertex" and "edge" to denote the overlapping node and the overlapping edge respectively. Let R be the overlapping number of cycles passing through an edge of G. We use $R_i$ to denote an edge of R=$i$ ($i$ is a natural number). In a cycle-set of G, a boundary vertex is a vertex being incident with two $R_1$ only, an interior vertex is a vertex being incident with $R_2$ only, and a cycle is removable implies that it has a unique $R_1$ whose two end vertices are boundary vertices. We use $C_x$ to denote a removable cycle such that the rest of the cycle-set is non-Hamiltonian when removing it from the cycle-set. A vertex that there are not less than 3 vertices of degree less than or equal to 2 in its neighborhood is denoted by $N_3$, $N_2$ denotes a vertex that there are 2 vertices of degree less than or equal to 2 in its neighborhood, and $N_1$ denotes a vertex that there are vertices of degree equal to 1. If there are no $N_3$, $N_2$ and $N_1$ in G, we say G is ($N_3$, $N_2$, $N_1$)-free. $R_2$-free denotes there are no $R_2$ in a cycle-set of G. Sometimes we combine $N_3$, $N_2$, $N_1$ and $R_2$ in one parentheses such as ($N_3$, $N_2$, $N_1$, $R_2$)-free. We use $F_s+C_x$ to denote recombination of $F_s$ and $C_x$ such that there are no other removable cycles except for $C_x$. Let ***F*** be a ($N_3$, $N_2$, $R_2$)-free subgraph of $F_s+C_x$. For terms and notations not defined here see [3] [4].

In this paper we investigate the relationship between $C_x$ and Hamilton cycles in a cycle-set of a graph. Generally speaking, for a given ($N_3$, $N_2$, $N_1$)-free subgraph of a Hamilton graph G, we could obtain a set of cycles by removing some cycles from it, and the union of these cycles yields a Hamilton cycle. But, for a graph that the Hamiltoncity is unknown, if there is a $C_x$ in the removed cycles, then the claim of Hamiltoncity of the given graph is sometimes not true (see **Lemma 2.3.**). Thus, we study the Hamiltoncity of edges in the structure of the cycle-set and develop a new method to determine the Hamiltoncity of the given graph. After excluding the graphs with $N_3$ and eliminating the paths which a Hamilton cycle must not pass through at $N_2$ and $N_1$ (see **Lemma 2.1.**, **Corollary 2.1.**), we obtain a ($N_3$, $N_2$, $N_1$)-free subgraph of G. For a cycle-set of this subgraph, by deleting all the removable cycles arbitrarily in a sequence one by one under the condition of $N_3$-free, we can detect $C_x$ out (see **Corollary 2.1., Lemma 2.2.**) and recombine $C_x$ and the subgraph. With the same method, we obtain the $F_s+C_x$ eventually. We show that a connected ***F*** is the necessary and sufficient condition to determine the given graph is Hamiltonian (see **Theorem 2.1., Theorem 2.2., Theorem 2.3.,** and **Corollary 2.2.**).

## 2. The proofs
**Lemma 2.1.** *Let G be a graph. For $N_2 \in V(G)$, the edge not being incident with the vertices of degree 1 or 2 in the neighborhood of $N_2$ is not a path which a Hamilton cycle must pass through.*

**Proof.** Let G be a graph. For $N_2 \in V(G)$, v denotes a vertex of degree 1 or 2 in the neighborhood of $N_2$. Since a Hamiltonian cycle (if has) will enter and leave each vertex in G exactly once, deleting just one edge of v only causes G non-Hamiltonian. In the other words, it means that all the edges not being incident with the vertices of degree 1 or 2 in the neighborhood of $N_2$ are not paths which a Hamilton cycle must pass through. □

**Corollary 2.1.** *Let G be a graph. For $N_3 \in V(G)$, G is non-Hamiltonian if it includes a $N_3$.*

**Proof.** Let $v_1, v_2, \ldots, v_i$ ($i \geq 3$) be the vertices of degree 1 or 2 in neighborhood of $N_3$ ($N_3 \in V(G)$). From **Lemma 2.1.**, since $v_1$ and $v_2$ constitute $N_2$ at vertex $N_3$, then the edge $N_2 v_i$ ($i \geq 3$) is not a path which a Hamilton cycle (if has) must pass through, and that implies no Hamilton cycles (if has) pass through $v_i$ ($i \geq 3$). Thus, G is non-Hamiltonian. □

**Theorem 2.1.** *$F_s$ is Hamiltonian if, and only if, its ($N_3$, $N_2$, $R_2$)-free subgraph is connected.*

**Proof.** For a spanning cycle-set $F_s$, induced from a given ($N_3$, $N_2$, $N_1$)-free graph G, we can divide the edges into three kinds: $R_1$, $R_2$, and $R_{3+i}$ (i is a natural number). By the definition, we know that no pair of cycles is same in a covering of a cycle-set over on the graph, therefore, $F_s$ never consists of only $R_2$ or $R_{3+i}$ and there have four cases as following: $\{R_1\}$, $\{R_1, R_2\}$, $\{R_1, R_{3+i}\}$, $\{R_1, R_2, R_{3+i}\}$. By the definition, an $F_s$ deleted any one of irremovable cycles is non-Hamiltonian, that implies a $R_1$ on the irremovable cycles is a path which a Hamilton cycle must pass through. On the other, we know that a $R_2$ is a public edge of tow cycles, if there are two $R_1$ on an endvertex of a $R_2$, then the $R_2$ is not a path which a Hamilton cycle must pass through by **Lemma 2.1..** Thus, the following analysis will base on the fact that a $R_1$ in an $F_s$ is irremovable and a $R_2$ is removable.

(i)$\{R_1\}$ If the edges in an $F_s$ consist of $R_1$ only, it is obviously that the edges in the ($N_3$, $N_2$, $R_2$)-free subgraph consist of $R_1$ only, too. If the subgraph is disconnected, then there must be a $R_2$ in $F_s$. It is clear that the case does not exist. And this means the subgraph is connected. Note that a $R_1$ in an $F_s$ is a path which a Hamilton cycle must pass through. Hence, the only result under the ($N_3$, $N_2$, $R_2$)-free condition is that the subgraph is a Hamilton cycle.

(ii)$\{R_1, R_2\}$ If the subgraph is disconnected then there must be a vertex of order 2 (as an interior vertex) on a $R_2$ in the $F_s$, and then the two $R_2$ are paths which a Hamilton cycle must pass through. But an $F_s$ is a graph without removable cycles, so there must be two $R_1$ being adjacent to the endvertex of an $R_2$. Therefore, the endvertex is a $N_3$. By **Corollary 2.1.**, the $F_s$ is non-Hamiltonian. Since an $F_s$ is a subgraph induced from a ($N_3$, $N_2$, $N_1$)-free graph, then, If the subgraph is connected then the Union of all elements in the ($N_3$, $N_2$, $R_2$)-free subgraph is actually a spanning cycle, that is the $F_s$ is Hamiltonian.

(iii)$\{R_1, R_{3+i}\}$ Without loss of generality, we consider the case that there are $3+i$ irremovable cycles on a $R_{3+i}$. Under ($N_3$, $N_2$, $R_2$)-free condition, by **Lemma 2.1.**, the two edges that are adjacent to each of the endvertices of the $R_{3+i}$ are only paths which a Hamilton cycle must pass through. So, there is only one result for the ($N_3$, $N_2$, $R_2$)-free subgraph, the subgraph is disconnected that means there are not enough paths in the $F_s$ to satisfy the existence of a Hamilton cycle.

(iv){$R_1, R_2, R_{3+i}$} We consider mainly the relationship between the location of a $R_2$ and the connectivity of the subgraph besides that the case of $R_1$ and $R_{3+i}$ is the same as (iii).

(1) In the case of that every $R_2$ is a chord of a spanning cycle, deleting a $R_2$ under ($N_3$, $N_2$, $R_2$)-free condition has no relationship with the connectivity of the subgraph. Hence, the case is the same as (iii).

(2) There are not less than two disjoint $R_2$ on a spanning cycle. If $R_2$ are deleted under ($N_3$, $N_2$, $R_2$)-free condition, then the subgraph of the $F_s$ is disconnected. Note that a $R_2$ in an $F_s$ is removable. Thus, it is obviously that there are not enough paths which a Hamilton cycle must pass through, that is the $F_s$ is non-Hamiltonian.

(3) If anyone of the $R_2$ is a chord of a cycle in $F_s$, then, the endvertices of the $R_2$ will be belong to two cycles respectively after deleting the $R_2$, it can only mean one thing that there is a cut vertex in the subgraph of the $F_s$. Clearly, the case is the same as (iii) also.

As proved above, a connected ($N_3$, $N_2$, $R_2$)-free subgraph is the necessary and sufficient condition of a Hamiltonian $F_s$. □

**Lemma 2.2.** *Let F be a cycles-set of ($N_3$, $N_2$)-free graph G. Every $R_1$ of $C_x$ in G is a path which a Hamilton cycle must pass through.*

**Proof.** By the definition of $N_2$ and $N_3$, we can easily construct a serious of non-Hamiltonian graphs by deleting one irremovable cycle such that all non-Hamiltonian graphs are isomorphic to one of these non-Hamiltonian graphs, showing in the Figure 1.

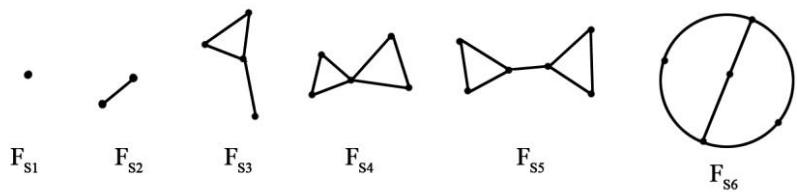

Figure 1

Therefore, we can construct a recombination of $F_s$ and $C_x$ (the irremovable cycle) showing in the Figure 2 below. It is clear that $F_s+C_x$ is Hamiltonian implies there exists a path which a Hamilton cycle must pass through in $C_x$. Since a $R_1$ is the only edge added into the graph (the doting line in the figure), then a $R_1$ of $C_x$ in G must be a path which a Hamilton cycle must pass through. □

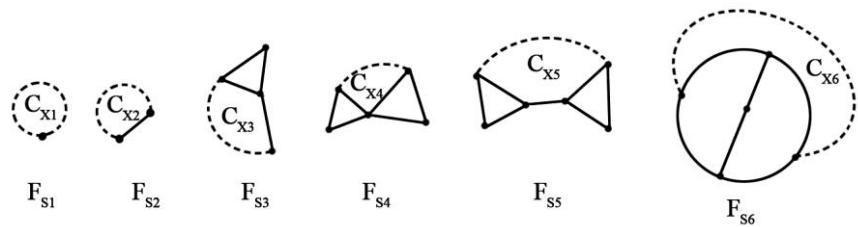

Figure 2

**Theorem 2.2.** $F_s+C_x$ is Hamiltonian if, and only if, **F** is connected.

**Proof.** Let G be a ($N_3$, $N_2$, $N_1$)-free graph, $F_s \subseteq G$, $C_x \subseteq G$, **F** is a ($N_3$, $N_2$, $R_2$)-free subgraph. We first prove a statement that no endvertex of $R_1$ in **F** is an isolated vertex. Since G is a ($N_3$, $N_2$, $N_1$)-free graph, then it is impossible for an endvertex existing as an isolated vertex in $F_s$ which induced from a cycle-set of G. Therefore, all $R_1$ is a path which a Hamilton cycle must pass through in **F** under the ($N_3$, $N_2$, $R_2$)-free condition induced from $F_s+C_x$, Hence, the statement is hold.

Since there have only $R_1$ and $R_{3+i}$ in **F**, and by **Theorem 2.1.** and **Lemma 2.2.**, all the $R_1$ in an $F_s$ or $C_x$ is a path which a Hamilton cycle must pass through, then it is enough to consider the case of $R_{3+i}$ in **F**. Know that there is no isolated vertex to be as an endvertex in **F**, thus only four cases of the $R_{3+i}$ existing with $R_1$ in **F** need considering as following (see the Figure 3),

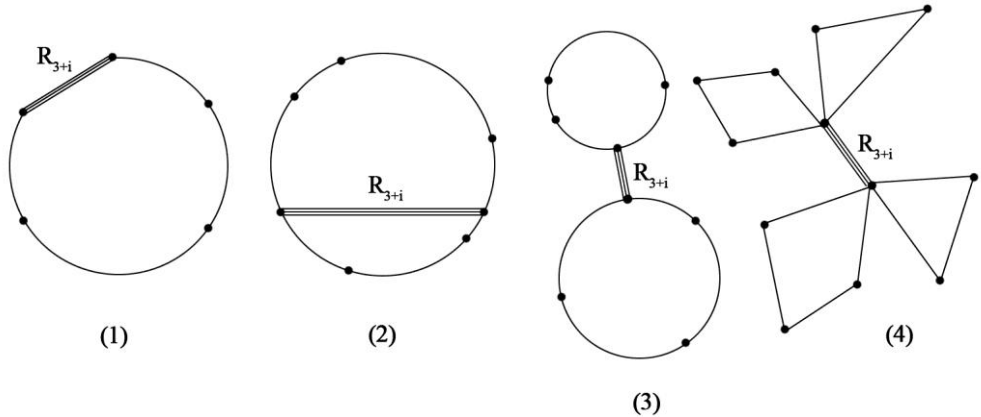

Figure 3

Case 1  $R_{3+i}$ is an edge of a cycle;
Case 2  $R_{3+i}$ is an chord of a cycle;
Case 3  $R_{3+i}$ is a bridge connected two cycles;
Case 4  $R_{3+i}$ is a bridge connected three or more cycles.

For Case 3 and Case 4, it is easy to determine, by **Lemma 2.1.**, that $R_{3+i}$ is a path which a Hamilton cycle must pass through so that it must be deleted when inducing **F** from $F_s+C_x$, thus, **F** is disconnected, and then $F_s+C_x$ is non-Hamiltonian clearly. For Case 2, similarly by **Lemma 2.1.**, $R_{3+i}$ must be deleted, so that **F** is connected, and then $F_s+C_x$ is Hamiltonian. For case 1, since no isolated vertex exists, then there is a cycle pass through every vertex once exactly, and then **F** is connected, $F_s+C_x$ is Hamiltonian. In summary, we derive that $F_s+C_x$ is Hamiltonian if, and only if, **F** is connected. Moreover, when adding a vertex (or more) of degree 2 into $R_{3+i}$, there is no change of the connectivity of **F** despite of stronger passing necessity to $R_{3+i}$ so that the conclusion above is hold. □

**Lemma 2.3.** *Let F be a cycle-set of ($N_3$, $N_2$)-free graph G. $C_r$ is all the irremovable cycles except $C_x$ ($C_x \in F$). $R_1$ in $C_r$ is a path by which a Hamilton cycle must pass in F.*

**Proof.** After deleting all irremovable cycles, if a spanning cycle-set induced from F is Hamiltonian, then we have two cases to consider in these irremovable cycles: $C_x=0$, $C_x \neq 0$. It is clear that $R_1$ in $C_r$ is a path which a Hamilton cycle must pass through in F no matter what case to consider. If a spanning cycle-set induced from F is non-Hamiltonian, we still have two cases: $C_x=0$, $C_x \neq 0$. By **Theorem 2.2.**, we know the only reason for $F_s+C_x$ is non-Hamiltonian is that a ($N_3$, $N_2$, $R_2$)-free edge induced subgraph from $F_s+C_x$ is disconnected, and there is no relationship to $R_1$ in $C_r$. Hence, the lemma is hold. □

**Theorem 2.3.** *Every two spanning cycle-sets in the given graph have the same Hamiltonian property with the given graph.*

**Proof.** Graph G is a ($N_3$, $N_2$)-free graph. Let $F_1$ and $F_2$ denote two different labeled cycle-sets in graph G. $F_{s1}$ and $F_{s2}$ denote the spanning cycle-set of $F_1$ and $F_2$ respectively. By the definition of a cycle-set, we know that the union of the cycles in a cycle-set of graph G is the graph G itself, so do $F_1$ and $F_2$. Therefore, there exist the same number of paths which a Hamilton cycle must pass through both in $F_1$ and $F_2$. Since every edge in G is correspondent to a value of R in an F, then we can determine the Hamiltoncity of an edge. By **Lemma 2.3.**, we know $R_1$ of $C_r$ is a path which a Hamilton cycle must not pass through, thus we can derive $F_{s1}$ and $F_{s2}$ by deleting all the irremovable cycles from G, and therefore, we can determine $R_2$ in Fs is a path which a Hamilton cycle must not pass through also. While by **Lemma 2.2.**, we know $R_1$ in Fs is a path which a Hamilton cycle must pass through. In addition, there is no change of the number of $R_{3+i}$ when inducing $F_s$ from F. So, all the paths which a Hamilton cycle must pass through in $F_1$ and $F_2$ remained in $F_{s1}$ and $F_{s2}$. Hence, $F_{s1}$ and $F_{s2}$ not only have same number of the paths in graph G but also have the same Hamiltoncity with graph G. That completes the proof. □

**Corollary 2.2.** *Graph G is Hamiltonian if, and only if, $F_s+C_x$ is Hamiltonian.*

**Proof.** By **Theorem 2.3.**, we know that there are no changes of the Hamiltoncity of a cycle-set (F) derived from anyone of the cycle-set of the given graph G, and so does every $F_s+C_x$ induced from graph G. □

References

[1] Ronald J. Gould, Advances on the Hamiltonian problem—a survey, Graphs Combinatorics，19 (1) (2003) 7-52.
[2] S.Goodman, S.Hedetniemi, Sufficent Condition for a graph to be Hamiltonian, Journal of Combinatorial Theory Ser. B 16 (1974) 175-180.
[3] Ronald J. Gould, Graph Theory, Dover Publications, Inc., New York, 2012.
[4] J.A.Bondy, U.S.R.Murty, Graph Theory with Applications, Fifth Printing, Elsevier Science Publishing Co., Inc., New York, 1982.